\documentclass[twocolumn]{jpsj2}
\def\be{\begin{equation}}
\def\ee{\end{equation}}
\def\beq{\begin{eqnarray}}
\def\eeq{\end{eqnarray}}

\def\runtitle{}
\def\runauthor{Masao Ogata}

\title{%
Superconducting States in Frustrating $t$-$J$ Model: A Model 
Connecting High-$T_c$ Cuprates, Organic Conductors and Na$_x$CoO$_2$}

\author{%
Masao {\sc Ogata}\thanks{E-mail address: ogata@phys.s.u-tokyo.ac.jp}
}

\inst{%
Department of Physics, University of Tokyo, \\
7-3-1, Hongo, Bunkyo-ku, Tokyo 113-0033 Japan}

\recdate{\today}

\abst{%
The two-dimensional $t$-$J$ model on a frustrating lattice is 
studied using mean-field variational theories with 
Gutzwiller approximation.
We find that a superconducting state with 
broken time-reversal symmetry (d+id state) is realized 
in the parameter region close to the triangular lattice.  
The frustration enlarges the region of superconductivity
when $t<0$ for the hole-doped case, which is equivalent to 
$t>0$ for electron doping. 
We also discuss the SU(2) degeneracy at half-filling.   
The d+id state probably corresponds to the 
spin gap state at half-filling. 
}

\kword{%
Frustration, $t$-$J$ model, time-reversal symmetry breaking, 
Gutzwiller approximation
}

\begin{document}
\sloppy
\maketitle

Frustration is an interesting problem and has been extensively 
studied in spin systems.  
The effect of frustration in itinerant electron systems, 
however, is not so well understood.
Resonating-valence bond (RVB) state, which is a liquid 
state consisting of spin-singlet bonds, was originally discussed 
in the Heisenberg model on a triangular lattice 
where the frustration was expected to destroy the long-range 
magnetic order leading to a quantum spin-liquid.\cite{Anderson,Fazekas}
It was speculated that the carrier doping to the spin system 
will induce the RVB states which are directly related to 
superconductivity.\cite{Anderson87}  
Thus it is very important to investigate where the RVB states 
are realized in frustrated systems.  

Recently it has become possible to dope carriers in the 
two-dimensional organic conductor, $\kappa$-(BEDT-TTF)$_2$$X$, 
which has a lattice structure very similar to the triangular lattice. 
Actually it shows superconductivity under pressure.\cite{Kanoda}
More recently, a superconductivity has been 
discovered in Na$_x$CoO$_2$ compounds, in which Co atoms form 
layered triangular lattices.\cite{Takada}  
Although the superconducting properties in these materials have 
not been clarified yet, it is very interesting to study the possibility 
of RVB-type superconductivity in a model which has the effect of 
frustration.  
In this paper we study the $t$-$J$ model on a lattice as 
shown in Fig.\ 1, which interpolates between 
the square and triangular lattices.  
The organic conductors have such lattice structures.  

In the case of high-$T_c$ superconductivity, hole doping to 
the square Heisenberg spin system induces d$_{x^2-y^2}$-wave 
RVB-type superconductivity.\cite{YSS,Gros,ZGRS,Tanamoto,YO}  
On the other hand, in the triangular lattice, the 
analysis of high-temperature expansion of the free energy 
suggests that the doping induces an RVB state when the 
hopping integral $t$ is positive for the hole-doped 
case.\cite{Koretsune}  
(Note that this corresponds to $t<0$ for electron-doped case.)
When $t<0$, the competition 
between the ferromagnetic tendency and singlet formation was 
observed.  
In this paper, 
in order to study the possibility of superconductivity in 
the ground state, 
we use a variational theory with Gutzwiller approximation.\cite{ZGRS,YO}  
We find that the frustration enlarges the region of superconductivity
when $t<0$, and that a superconducting state with 
broken time-reversal symmetry (chiral state) is realized 
in the parameter region close to the triangular lattice.  
We also discuss the SU(2) degeneracy at half-filling.   

\begin{figure}[t]
\begin{center}
\includegraphics[width=5cm]{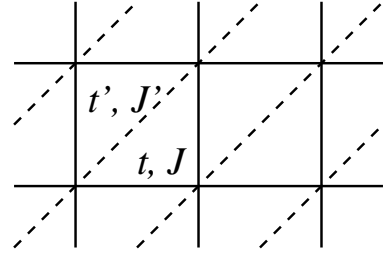}
\end{center}
\caption{Lattice structure of the frustrated two-dimensional 
$t$-$J$ model.
On the bonds forming the square lattice (solid lines), we have 
$t$ and $J$ terms.  $t'$ and $J'$ terms are defined 
on the diagonal bonds (dashed lines).
\label{fig:1}}
\end{figure}

The $t$-$J$ model on a lattice in Fig.\ 1 is written as
\begin{align}
{\cal H}= &-t \sum_{(i,j)\sigma}P_G(c_{i\sigma}^\dagger c_{j\sigma}
+{\rm H.c.}) 
P_G + J\sum_{(i,j)} {\mib S}_i \cdot {\mib S}_j \cr
&-t' \sum_{\langle i,j \rangle\sigma}P_G(c_{i\sigma}^\dagger c_{j\sigma}
+{\rm H.c.}) 
P_G + J'\sum_{\langle i,j \rangle} {\mib S}_i \cdot {\mib S}_j,
\end{align}
where $P_G= \prod_i (1-n_{i\uparrow}n_{i\downarrow})$ is the 
Gutzwiller projection operator which excludes double occupancies. 
$(i,j)$ and $\langle i,j \rangle$ represent the summation over 
nearest-neighbor and one of the next-nearest neighbor pairs 
as shown in Fig.\ 1, respectively. 
The sign of $t$ is crucial since there is no electron-hole 
symmetry.\cite{Koretsune}
Hereafter we use $|t|$ as an energy unit and study 
the parameter range, $t'/t=J'/J=0.0$-$1.0$.
The model becomes the square lattice $t$-$J$ model when $J'/J=0$;
and the triangular lattce when $J'/J=1$.

For this Hamiltonian, we consider Gutzwiller-type variational 
wave function $P_G |\Phi \rangle$ 
with $|\Phi\rangle$ representing a BCS mean-field wave function.  
For estimating the variational energy, we use 
the Gutzwiller approximation in which the 
effect of the projection is replaced by a statistical weight 
as follows\cite{ZGRS,YO}
\begin{align}
 \langle c^\dagger_i c_j \rangle &= g_t \langle c^\dagger_i c_j \rangle_{0},  
\nonumber\\
 \langle {\mib S}_i \cdot {\mib S}_j \rangle &= 
g_s \langle {\mib S}_i \cdot {\mib S}_j \rangle_{0}.
\end{align}
Here $\langle \cdots \rangle_0$ represents the expectation value 
in terms of $|\Phi\rangle$, 
and $\langle \cdots \rangle$ represents the normalized expectation 
value in $P_G|\Phi\rangle$.   
In the simplest Gutzwiller approximation we have 
\be
g_t = \frac{2\delta}{1+\delta}, \qquad 
g_s = \frac{4}{(1+\delta)^2},
\ee
where $\delta$ is the density of holes, i.e., the electron 
number is $n=1-\delta$.  
It has been shown that this approximation gives a fairly 
reliable estimation for the variational energy for the 
pure d$_{x^2-y^2}$-wave superconducting state when it is 
compared with the variational Monte Carlo results.\cite{ZGRS,YO}  

In order to search for the state with the lowest variational energy, 
we studied large unit cells assuming site dependent order parameters.
In this Gutzwiller approximation, we can have order parameters, 
\begin{equation}
\Delta_{ij} = 
\langle c_{i\uparrow}^\dagger c_{j\downarrow}^\dagger \rangle_0, 
\quad 
\chi_{ij\sigma} = \langle c_{i\sigma}^\dagger c_{j\sigma} \rangle_0.  
\end{equation}
We assume $\chi_{ij\uparrow}=\chi_{ij\downarrow}=\chi_{ij}$ 
and that $\Delta_{ij}$ is a singlet pairing, i.e., 
$\Delta_{ij}=\Delta_{ji}$.  
By minimizing the variational energy we obtain 
a Bogoliubov-de Gennes equation\cite{Himeda}
and self-consistency equations.  
We have tried various trial states and found that the uniform 
state is always stabilized except for  
some special states realized only very near half-filling. 
Since these special states may be an
artifact of the mean-field-type calculations,\cite{Ogata} 
they are not discussed in this paper.  
The obtained phase diagrams for $J/t=0.3, J'/t'=0.3$ are 
shown in Fig.\ 2 as a function of the ratio $J'/J$ and the 
electron density $n$.


\begin{figure}[t]
\begin{center}
\begin{tabular}{c}
   \resizebox{80mm}{!}{\includegraphics{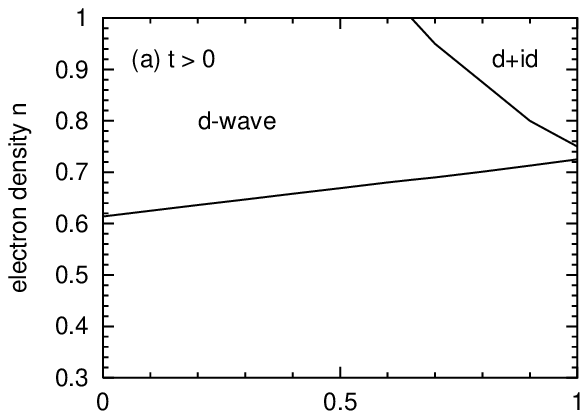}} \\
   \resizebox{80mm}{!}{\includegraphics{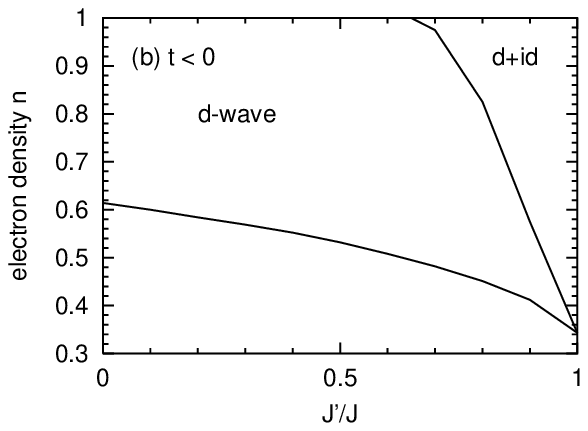}} 
\end{tabular}
\end{center}
\caption{Phase diagram obtained in Gutzwiller approximation 
of the RVB states in the frustrated two-dimensional $t$-$J$
model with (a) positive $t$ and (b) negative $t$. $J/|t|=0.3$
and $J'/|t'|=0.3$.
\label{fig:2}}
\end{figure}

We find that in the region with $J'/J<0.675$ the d$_{x^2-y^2}$-wave 
superconducting state is stabilized.  This state is the same as in 
the square lattice case ($J'=0$).  
When $J'/J$ exceeds 0.675 a superconducting state with imaginary 
part of order parameter appears, 
which we call as d+id state as shown in Fig.\ 2.  
Since there is no continuous symmetry in our model, the notation 
of d+id-wave just represents a discretized version of 
d$_{x^2-y^2}$-id$_{xy}$-wave superconductivity.  
The detailed analysis is shown later.
This state has excitation gaps all over the 
fermi surface and breaks time reversal symmetry.  
Another interesting feature is that for $t<0$ the superconducting 
region increases as $J'/J$ becomes large, while its region decreases 
when $t>0$.  

Before discussing the superconducting properties, let us here 
clarify the SU(2) degeneracy realized at half-filling in the 
present model.
By solving the Bogoliubov-de Gennes equations we find 
various states at half-filling which have exactly the same 
variational energy with each other.  
For the uniform state, we have the energy dispersion for quasiparticles 
as 
\be
E_k = \sqrt{\varepsilon_k^2 + |F_k|^2}, 
\ee
where
\begin{align}
\varepsilon_k &= - \frac{3}{2}g_s J (\chi_x \cos k_x + \chi_y \cos k_y)
- \frac{3}{2}g_s J' \chi_{xy} \cos(k_x+k_y), \\
F_k &=  \frac{3}{2}g_s J (\Delta_x {\rm e}^{-i\theta}\cos k_x + 
\Delta_y {\rm e}^{i\theta'}\cos k_y)
+ \frac{3}{2}g_s J' \Delta_{xy} \cos(k_x+k_y). 
\end{align}
Here $\Delta_x$ ($\Delta_y$) is the superconducting order 
parameter on the bonds in the $x$-($y$-)direction, and 
$\Delta_{xy}$ is in the diagonal direction.   
We denote $\theta$ ($\theta'$) as the phase of the order 
parameter relative to that on the diagonal bonds.
Note that, in this notation, the pure d$_{x^2-y^2}$-wave 
state is represented as 
$\Delta_x=\Delta_y$, $\Delta_{xy}=0$ and $\theta=\theta'=\pi/2$.

We find that, in each energetically 
degenerate state, these parameters satisfy the following relations;
\begin{subequations}
\begin{align}
&\chi_x \chi_y + \Delta_x \Delta_y \cos (\theta + \theta') =0 ,\\
&\chi_x \chi_{xy} + \Delta_x \Delta_{xy} \cos \theta =0 ,\\
&\chi_y \chi_{xy} + \Delta_y \Delta_{xy} \cos \theta' =0 .
\end{align}
\end{subequations}
In this case the energy dispersion becomes 
\be
E_k = \frac{3}{2}g_s \sqrt{
J^2(\chi_x^2+\Delta_x^2)\cos^2 k_x + 
J^2(\chi_y^2+\Delta_y^2)\cos^2 k_y + 
J'^2(\chi_{xy}^2+\Delta_{xy}^2)\cos^2 (k_x + k_y)},
\ee
The self-consistency equations are 
\begin{subequations}
\begin{align}
1 &=\frac{1}{N}\sum_{k}\frac{\frac{3}{4}g_s J \cos^2 k_x}{E_k}, \\
1 &=\frac{1}{N}\sum_{k}\frac{\frac{3}{4}g_s J \cos^2 k_y}{E_k}, \\
1 &=\frac{1}{N}\sum_{k}\frac{\frac{3}{4}g_s J' \cos^2 (k_x+k_y)}{E_k},  
\end{align}
\end{subequations}
where $N$ is the number of sites and the summation over 
$\mib{k}$ is in the Brillouin zone (in our model $-\pi \le 
k_x, k_y \le \pi$).

Since there are 8 parameters ($\chi_x, \chi_y, \chi_{xy}, \Delta_x, 
\Delta_y, \Delta_{xy}, \theta, \theta'$) and 6 equations
(8) and (10), we 
have remaining two degrees of freedom which lead to the 
various states obtained at half-filling.  
This is an extension of the SU(2) degeneracy discussed in the 
square lattice ($J'=0$).\cite{ZGRS,Affleck}  
For example, one of the realized states at half-filling has
\be
\begin{split}
\chi_x=\chi_y=0, \quad \chi_{xy} \ne 0, \quad
&\Delta_x=\Delta_y \ne 0, \quad \Delta_{xy}=0, \\
\theta=\theta'&=\frac{\pi}{4}.
\end{split}
\ee
When $\chi_{xy}\rightarrow 0$, this state is naturally connected to the 
so-called d+is state realized in the square lattice.  
Another state with 
\be
\begin{split}
\chi_x=\chi_y=\Delta_x &=\Delta_y \ne 0, \\
\chi_{xy}=0, \quad \Delta_{xy}\ne 0 \ &{\rm and\ }
\theta=\theta'=\frac{\pi}{2},
\end{split}
\ee
is connected to the d$_{x^2-y^2}$-wave state for the square lattice.  

Since these degenerate states have common values for
\begin{align}
D &= \sqrt{\chi_x^2 + \Delta_x^2}= \sqrt{\chi_y^2 + \Delta_y^2}, \\
D' &= \sqrt{\chi_{xy}^2 + \Delta_{xy}^2},
\end{align}
we plot their $J'/J$ dependence in Fig.\ 3(a).  
As $J'/J$ approaches to 1, $D$ becomes equal to $D'$.  

\begin{figure}[t]
\begin{center}
\begin{tabular}{c}
   \resizebox{80mm}{!}{\includegraphics{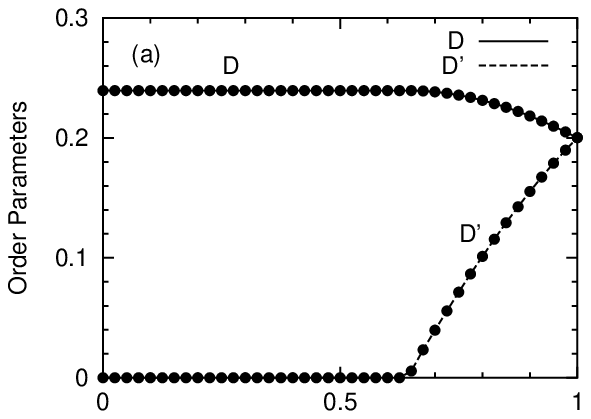}} \\
   \resizebox{80mm}{!}{\includegraphics{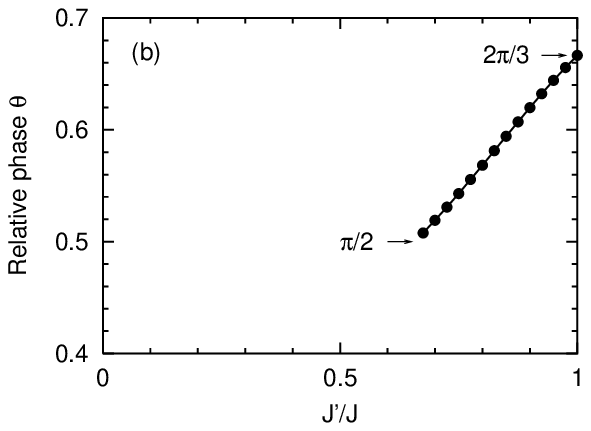}} 
\end{tabular}
\end{center}
\caption{$J'/J$ dependences of (a) order parameters and (b) relative 
phase $\theta=\theta'$, at half-filling.  See details in the text.
\label{fig:3}}
\end{figure}

Next let us discuss the doped case.  Once some holes are introduced, 
the above degeneracy is readily lifted, and the lowest non-degenerate 
state becomes the superdoncuting state which is shown in Fig.\ 2.
We note that the flux phase proposed before\cite{LeeFeng,WWZ,KL,ASH,} 
is not stabilized. 
Since the order parameters are uniquely determined at finite 
$\delta$, we determine the relative phase of the superconducting 
order parameters, $\theta$ and $\theta'$, 
by taking the limit $\delta\rightarrow 0$. 
The obtained phase is shown in Fig.\ 3(b). 
When $\Delta_{xy}$ starts to have a finite value at $J'/J=0.675$, 
the relative phases are $\theta=\theta'=\pi/2$. 
This means that a small component $\Delta_{xy}$ appears 
as an imaginary part (i$\Delta_{xy}$) 
in addition to the pure d$_{x^2-y^2}$-wave state.   
As $J'/J$ increases, the relative phase approaches to $2\pi/3$, and 
when $J'=J$ (triangular lattice) the three order parameters 
$\Delta_x, \Delta_y$ and $\Delta_{xy}$ become symmetric with 
each other.  This state is consistent with the other 
works.\cite{LeeFeng,footnote} 
This gradual change of the relative phase gives a 
natural continuation between the d$_{x^2-y^2}$-wave superconductivity 
in the square lattice and 120-degree phase in the triangular lattice.

In Fig.\ 4 we show the doping dependences of the order 
parameters $\Delta=\Delta_x=\Delta_y=\Delta_{xy}$ and 
$\chi=\chi_x=\chi_y=\chi_{xy}$ determined self-consistently 
at $t'=t, J'=J$ and $J/t=0.3$ (triangular lattice).
Qualitatively similar figure has been obtained recently,\cite{footnote}
but the order parameter $\chi$ has not been taken into account, 
which plays an important role in the gauge theories for the $t$-$J$ model. 
As shown in Fig.\ 4, the superconducting region is large, 
which is consistent 
with the high-temperature expansion study for the triangular 
$t$-$J$ model when $t>0$ .  
Surprisingly the superconducting region is even 
larger for $t<0$, in which case the high-temperature study has shown 
that the ferromagnetic instability takes place 
for $J/t<0.3$.\cite{Koretsune2}
In order to clarify the actual ground state in the triangular $t$-$J$ 
model, more careful energy comparisons including ferromagnetism as well
as the 120-degrees antiferromagnetism (realized at half-filling) are 
necessary.  
In particular, there will be a competition between the ferromagnetic 
fluctuation and singlet formations for $J/t>0.3$ which leads to a 
heavy fermion-like behavior appearing in the specific
heat.\cite{Koretsune}
The relation to this competition and the RVB states remains a 
future problem.  
In the region with very low electron density, we expect a 
triplet superconductivity.

\begin{figure}[t]
\begin{center}
\begin{tabular}{c}
   \resizebox{80mm}{!}{\includegraphics{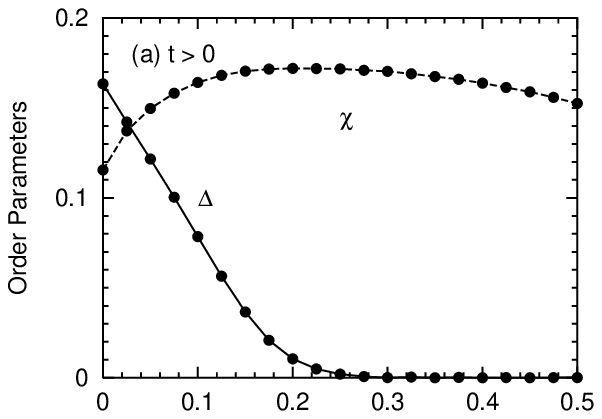}} \\
   \resizebox{80mm}{!}{\includegraphics{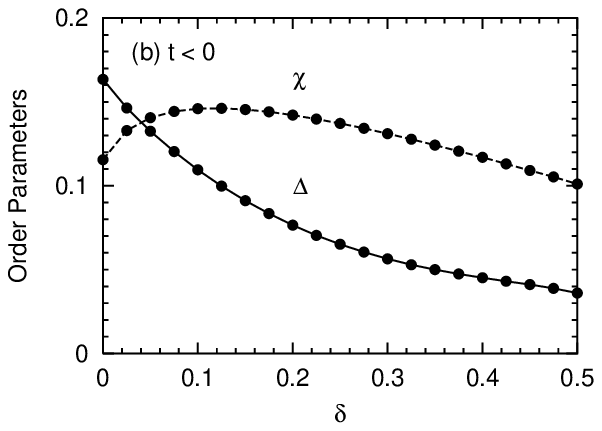}} 
\end{tabular}
\end{center}
\caption{Doping dependences of the order parameters in the triangular $t$-$J$
model with (a) positive $t$ and (b) negative $t$. $J/|t|=0.3$.
\label{fig:3}}
\end{figure}

For the recently discovered superconductor, Na$_x$CoO$_4$, the 
sign of $t$ is controversial.\cite{footnote,Singh}
If the sign is negative, the superconductivity is more favored.
For the organic superconductors, the sign of $t$ is negative, 
but in this case, the electron density is just at half-filling.  
In the $t$-$J$ model calculation, we cannot discuss this 
superconductivity and instead the Hubbard model has been used 
to explain the superconductivity via 
spin-fluctuation.\cite{Kino} 
In such calculations, only d$_{x^2-y^2}$-wave order parameter 
is observed.  It will be an interesting problem to search for 
the d+id state in the parameter close to the triangular lattice 
in the spin-fluctuation mechanism. 

There is another competition between antiferromagnetism and 
superconductivity in the phase diagram shown in Fig.\ 2.  
For the Heisenberg model, Weihong et al. discussed 
that the commensurate antiferromagnetism is stable 
for $J'/J<0.7$ using 
series expansion methods.\cite{Weihong} 
This critical value is the same as found in Fig.\ 2, which 
means that the d+id state probably corresponds to the 
spin gap state at half-filling. 
It is necessary to compare the energies of antiferromagnetic 
state and d-wave superconductivity in the phase diagram.  
However it is well known that the mean-field theories 
overestimate the magnetic states.  The Gutzwiller approximation 
does not remedy this problem.\cite{ZGRS}
In the square lattice, the variational Monte Carlo 
studies have shown that the lowest 
variational state is a coexistent state of antiferromagnetism 
and d-wave superconductivity for $\delta<0.1$.\cite{Giamarchi,Himeda2,OH}
As $J'/J$ increases and the magnetic frustration is induced, 
we expect that the region of antiferromagnetism shrinks and 
is confined in the very vicinity of half-filling.  
Therefore the RVB superconducting region in Fig.\ 2 will 
safely survive.  
Although the detailed study on this issue is necessary, 
the phase diagram shown in Figs.\ 2
and 4 promises that an interesting time-reversal breaking 
superconducting state is one of the strong candidates for the ground 
state.

In summary we have studied the frustrating $t$-$J$ model within 
a Gutzwiller approximation. 
The frustration induces a superconductivity with 
broken time-reversal symmetry 
in the parameter region close to the triangular lattice.  
The relative phase between the order parameters gradually
changes as a function of $J'/J$ which connects the 
d$_{x^2-y^2}$-wave and typical 120-degree phase in the 
triangular lattice.
We also discussed the SU(2) degeneracy at half-filling.   
We found that there is unexpectedly very large variety 
of states in strongly correlated frustrating systems.

The author would like to thank very useful discussions and 
conversations with B.\ S.\ Shastry, G.\ Baskaran, and 
T.\ Koretsune.

\end{document}